# Dynamic FMR and magneto-optical response of hydrogenated FCC phase $Fe_{25}Pd_{75}$ thin films and micro patterned devices

Shahbaz Khan[1,2], Satyajit Sarkar[3,6], Nicolas B. Lawler[1,4], Ali Akbar[5], Muhammad Sabieh Anwar[5], Mariusz Martyniuk[6], K. Swaminathan Iyer[5] and Mikhail Kostylev[1]

[1] Department of Physics and Astrophysics, The University of Western Australia, 35 Stirling Highway, 6009, Crawley, Perth, WA, Australia

[2] Commonwealth Scientific and Industrial Research Organisation (CSIRO), 26 Dick Perry Avenue, Kensington, WA 6151, Australia

[3] Department of Mechanical Engineering, The University of Western Australia, 35 Stirling Highway, 6009, Crawley, Perth, WA, Australia

[4] School of Chemistry and Biochemistry, The University of Western Australia, 35 Stirling Highway, 6009, Crawley, Perth, WA, Australia

[5] Department of Physics, Syed Babar Ali School of Science and Engineering, Lahore University of Management Science, Opposite Sector U, 54792, D.H.A. Lahore, Pakistan

[6] Department of Electrical, Electronics & Computer Engineering, The University of Western Australia, 35 Stirling Highway, 6009, Crawley, Perth, WA, Australia

# Abstract

In this work, we investigate the effects of $H_2$ on the physical properties of $Fe_{25}Pd_{75}$. Broadband ferromagnetic resonance (FMR) spectroscopy revealed a significant FMR peak shift induced by $H_2$ absorption for the FCC phased $Fe_{25}Pd_{75}$. The peak shifted towards higher applied fields, which is contrary to what was previously observed for CoPd alloys. Additionally, we conducted structural and magneto-optical Kerr ellipsometric studies on the $Fe_{25}Pd_{75}$ film and performed density functional theory calculations to explore the electronic and magnetic properties in both hydrogenated and dehydrogenated states. In the final part of this study, we deposited a $Fe_{25}Pd_{75}$ layer on top of a microscopic coplanar transmission line and investigated the FMR response of the layer while driven by a microwave current in the coplanar line. We observed a large amplitude FMR response upon hydrogen absorption, as well as desorption rates when cycling between pure $N_2$ and a mixture of 3% $H_2$ + 97% $N_2$.

# Introduction

Renewable energy technologies have recently garnered significant interest as potential replacements for conventional, fossil fuel-based technologies. Hydrogen-based energy technologies, including solid-state hydrogen storage and energy conversion devices such as fuel cells, are currently a popular research topic and have potential for large-scale commercial applications. Palladium (Pd) is a unique material with a strong affinity for hydrogen due to its catalytic and hydrogen-absorbing properties. Pd has the potential to play a significant role in nearly every aspect of the envisioned hydrogen economy [1], including hydrogen purification [2], storage [3], detection [4-6], and fuel cells [7]. As $H_2$ molecules approach the surface of Pd, Pd's catalytic effect dissociates the $H_2$ molecule into two H atoms [8], which can then quickly diffuse into the interspace of the Pd lattice, causing it to expand. This expansion alters the lattice spacing of the crystal structure, subsequently modifying Pd's electronic properties, changing the electron density of states, and affecting its optical properties. Chang et al. and Leung et al. have reported potential applications for $H_2$ gas sensing using Pd's spintronic properties through FMR [4, 9]. Changes were observed in ferromagnetic-nonmagnetic metal (FM-NM) systems based on CoPd layered films. These studies demonstrate the potential application of FM-NM systems in detecting $H_2$ gas.

However, the FM-NM multilayer technology has some drawbacks concerning the sensor's response time, FMR signal strength, and the need for sophisticated electronics as part of the FMR response detection mechanism, which hinders its commercialization potential. An alternative approach is to use materials that produce a significant FMR signal, regardless of the film's thickness, such as CoPd alloy [10, 11]. Subsequently, researchers also demonstrated that the magnetism of FePd alloy films is sensitive to the presence of hydrogen gas in the environment [10]. The effect of hydrogen on this composition was reported to strongly depend on the Pd atomic ratio in the Fe matrix [10].

A straightforward method to investigate changes in the magnetic properties of an FM-NM alloy or multilayer induced by the presence of hydrogen gas is FMR [4]. This paper presents a comprehensive study on the impact of hydrogen gas on the magnetic properties of FePd alloy films grown on silicon substrates, focusing on the FMR response. We demonstrate that one alloy composition, FCC $Fe_{25}Pd_{75}$ (designated as $Fe_{25}Pd_{75}$), exhibits a substantial change in FMR response when exposed to hydrogen gas. In our initial findings, we observed a distinctive FMR peak shift in the $Fe_{25}Pd_{75}$ alloy when exposed to hydrogen. This behavior, which is central to the aims of our study, was not as pronounced in the other alloys we examined. This investigation is supported by structural characterization of the alloy, magneto-optical Kerr effect characterization in the presence of hydrogen gas, and density functional theory-based simulations. Furthermore, we fabricate and evaluate a simple device prototype of an FMR-based hydrogen gas sensor employing an FePd film grown on the surface of a microscopic microwave microstrip line made of gold and aluminum contacts (Au-Al).

## Methods

Several FePd alloy films with varying Pd contents were deposited on undoped (111) silicon substrates using co-sputtering from Fe and Pd targets, employing a biased target sputtering instrument from 4'Wave Inc. The base pressure was maintained at $9.3 \times 10^{-9}$ torr. Argon, used as the working gas, had a flow rate of 35 sccm, and its pressure was maintained at $6 \times 10^{-5}$ torr during the deposition. Initially, the targets were pre-sputtered for 20 minutes to eliminate potential surface contamination before film deposition. The FePd films were co-sputtered from two 99.99 wt.% purity targets, purchased from Kurt J. Lesker. Square-shaped pulses were used to apply a -700 V target bias voltage for deposition, with a duty cycle frequency of 10 kHz (100 μs per cycle). Film stoichiometry was controlled by adjusting the on-time pulse width for the targets. For example, to deposit $Fe_{25}Pd_{75}$ films, our primary interest, 80 μs long pulses were applied to the Pd target and 20 μs ones to the Fe. To deposit films with other compositions, the on-times were adjusted proportionally. The substrate stage was maintained at 20 °C with forced water cooling and rotated at 20 revolutions per minute during film deposition to achieve better uniformity and edge coverage. Film thickness was chosen in the 30-35 nm range, which is large enough for the samples to exhibit a strong FMR signal. In total, four types of FePd alloy thin films were fabricated with different compositions – $Fe_{15}Pd_{85}$, $Fe_{25}Pd_{75}$, $Fe_{35}Pd_{65}$, and $Fe_{80}Pd_{20}$. However, we observed a significant FMR peak shift in the presence of $H_2$ only in the $Fe_{25}Pd_{75}$ alloy. Consequently, we focused our resources and time exclusively on the $Fe_{25}Pd_{75}$ alloy. Furthermore, we did not observe any significant effects of oxidation on our films, within a year of its fabrication during which this research was conducted. Hence, we chose not to use any capping layer.

In order to evaluate the potential of the films for hydrogen gas sensing applications, we performed FMR characterization in the presence of hydrogen gas. For this purpose, a film was placed into a custom-built gas-tight chamber, the floor of which consists of a 0.4 mm wide microwave microstrip line. The sample was placed on top of the microstrip line, with the film facing the microstrip line. A constant frequency microwave signal was applied to the microstrip line from a microwave generator. The resulting microwave current passing through the microstrip line produced a microwave Oersted field around it, which induced magnetization dynamics within the FePd alloy film. We also applied a DC magnetic field perpendicular to the film, placing the sample between the poles of an electromagnet to create conditions favorable for an FMR response from the film. We recorded the FMR absorption peak by sweeping the DC magnetic field "applied field" from 0 to 10,000 Oe and monitored any shift in its position when exposed to $H_2$ gas. We repeated this process for multiple microwave frequencies, ranging from 2 to 18 GHz.

The FMR absorption information is contained within the amplitude of the microwave signal after it has traversed the sample under investigation. At resonance, the amount of microwave power absorbed by the sample increases, leading to a decrease in the amplitude of the microwave signal at the output of the microstrip line. To register the FMR peak, the signal from the output port of the line is fed into a microwave diode. The DC output of the diode is then applied to a digital lock-in amplifier. To enable the lock-in detection of the FMR absorption peak, a small AC magnetic field at a frequency of 220 Hz is superimposed on top of the DC bias field. The AC and DC fields are maintained parallel. This modulating field is generated by a small coil, which is powered by an AC current produced by a function generator. The lock-in is referenced by the function-generator signal. This substantially enhances the setup's sensitivity; however, there is a technical detail associated with this method. The lock-in output signal represents the first derivative of the absorption peak with respect to the applied field, rather than the resonance absorption peak itself. This type of FMR trace is referred to as "a differential FMR absorption trace" [12].

To carry out the measurements in the presence of hydrogen gas, the hermetic chamber was filled and continuously fed with hydrogen gas at a flow rate of 500 standard cubic centimeters per minute (sccm). Prior to hydrogen exposure, baseline FMR traces were acquired in the sample's pristine state. These control measurements were conducted with 100% nitrogen gas, serving as a surrogate for air. Nitrogen was continuously streamed into the chamber at an identical flow rate of 500 sccm. This setup allowed for dynamic switching between $N_2$ and $N_2+H_2$ environments, thereby facilitating the repeated execution of FMR measurements under varying conditions.

XRD spectra were recorded using a Panalytical Empyrean XRD system, and the measurements were taken in Bragg Brentano (BB) geometry. Initially, the samples were placed on a 5-axis stage, and the spectra were recorded in a nitrogen environment. Subsequently, another measurement was taken in the same configuration in the hydrogen gas environment, followed by another trace in the same nitrogen environment.

High-resolution transmission electron microscopy (TEM) and Selected Area Electron Diffraction (SAED) images of the $Fe_{25}Pd_{75}$ sample were taken with Titan G2 80-200 TEM at 200 kV and 0.2 nm probe current. All measurements were performed in the same environment and under the same operational conditions. To prepare the TEM sample, a section smaller than 100 nm was cut from a large area of the thin film and attached to a special TEM grid with the aid of a focused ion beam (FIB). The TEM grid was then attached to a TEM holder to be brought into the TEM for further characterization.

Magneto-Optical Kerr effect (MOKE) on the thin films was performed in longitudinal geometry. A photoelastic-modulator (PEM) based technique was used, allowing the application of phase-sensitive detection methodology employing a lock-in amplifier. The sample was mounted on a flat surface inside a closed chamber with a thin quartz window. The sample holder was introduced within the pole pieces of an electromagnet. The sample surface was parallel to the electromagnet pole pieces, ensuring that the applied magnetic field was perpendicular to the sample surface as required for longitudinal MOKE geometry. A He-Ne laser ($\lambda$= 632.8 nm) with linear polarization (s polarization) was incident on the sample at an angle of 30° to the sample normal. The reflected intensity was detected with a photodetector, and its amplified response was measured using a lock-in amplifier. The PEM served as the lock-in reference.

Density functional theory (DFT) calculations were performed using Korringa-Kohn-Rostoker (KKR) code (with a coherent potential approximation) to explore the effect of atomic hydrogen on the structural, electronic, and magnetic properties of $Fe_{25}Pd_{75}$ surfaces [13, 14]. The generalized gradient approximation by Perdew, Burke, and Ernzerhof (GGA-PBE) was used to address the electron-ion interactions [15]. The system was optimized for $Fe_{25}Pd_{75}$ grown in the [111] direction, and the lattice parameter and bond length were determined for a total energy-minimum fitting using the Birch-Murnaghan equation of state before starting the DOS calculations [16, 17]. Thereafter, the $Fe_{25}Pd_{75}$ structure was introduced to a single hydrogen atom by considering all the available binding sites. The hydrogen atom was assumed to sit at the octahedral interstitial site of the $Fe_{25}Pd_{75}$ FCC lattice. For electronic properties, the Brillouin zone was set under the Monkhorst-Pack scheme with k-points [18], and the binding energy of hydrogen with $Fe_{25}Pd_{75}$ was calculated.

# Results and discussion

### FMR of the fabricated *samples*

Fig. 1 displays the raw differential FMR absorption traces for the fabricated samples with different compositions. All fabricated samples, except $Fe_{15}Pd_{85}$, demonstrated an FMR response peak. From Fig. 1 it is evident that all the films are responding to the presence of 100% hydrogen gas in the film environment. The sample with the $Fe_{25}Pd_{75}$ stoichiometry ratio exhibits the most significant FMR peak shift upon the introduction of hydrogen gas into the sample environment. Consequently, our focus in the following study is solely on this chemical composition. It is worth noting that an earlier study [10] demonstrated that the absorption of hydrogen gas alters the magnetization of FePd samples with different Fe contents – $Fe_{40}Pd_{60}$ and $Fe_{30}Pd_{70}$. However, those measurements were taken with MOKE, not FMR. Furthermore, we are not aware of earlier FMR measurements of $Fe_{25}Pd_{75}$ films, and no reports were found on detailed studies of films with this stoichiometry. Therefore, in the following study, we report a detailed structural and magnetic characterization of the $Fe_{25}Pd_{75}$ film.

Fig. 2(a) displays the Bragg-Brentano (BB-XRD, Cu-Kα radiation at 40 kV and 40 mA) patterns of the $Fe_{25}Pd75$ alloy film in its as-deposited state (red line), in the presence of 3% $H_2$ in nitrogen as a carrier gas ("hydrogenated", black line), and after the evacuation of the $H_2/N_2$ gas mixture from the sample environment (green line). Analysis of the X-ray diffraction patterns reveals that the growth of the film plane is parallel to only the (111) and (222) plane peaks. This indicates that the alloy film has an FCC crystal structure with a perfect [111] fiber texture in the film's normal direction. Such fiber texture in thin metallic films deposited using the biased target ion beam deposition technique has been reported previously [19, 20].

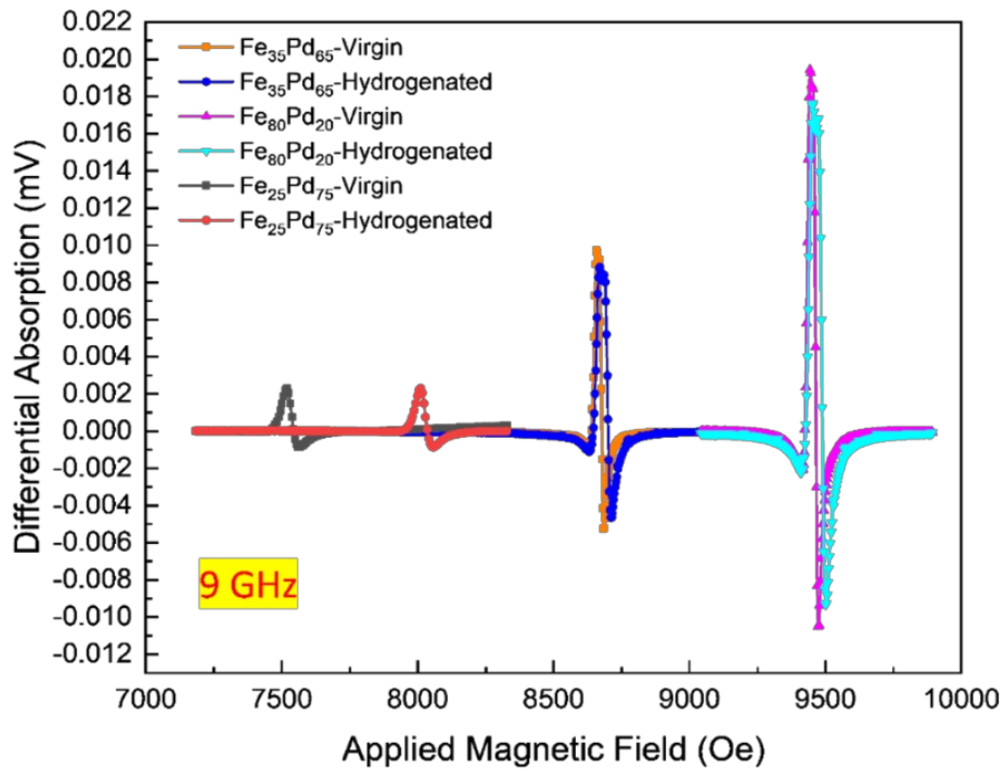

*Fig. 1: Shows the FMR response of deposited FePd alloy films in their virgin state and in the presence of 100% hydrogen gas as a function of alloy compositions. The measurements were taken at a microwave frequency of 9 GHz.*

Moreover, the diffraction peaks of the hydrogenated film are positioned at lower 2θ angles compared to the state of the film before hydrogenation. Two potential factors might contribute to similar peak shifts toward lower diffraction angles: (i) lattice dilation and (ii) in-plane compressive stresses [19]. Lattice dilation suggests an increase in lattice dimensions, possibly due to the incorporation of hydrogen ions/atoms into the metallic film matrix [21].

Additionally, the introduction of hydrogen may also induce compressive stress in the film in the in-plane direction, which could result in lattice expansion in the film's normal (out-of-plane) direction, as evidenced by the corresponding XRD peak shifts upon hydrogenation. Subsequently, we also observed changes in the magnetic properties of the thin films with the introduction of hydrogen in the thin film matrix, as described by the MOKE hysteresis loops in Fig. 4.

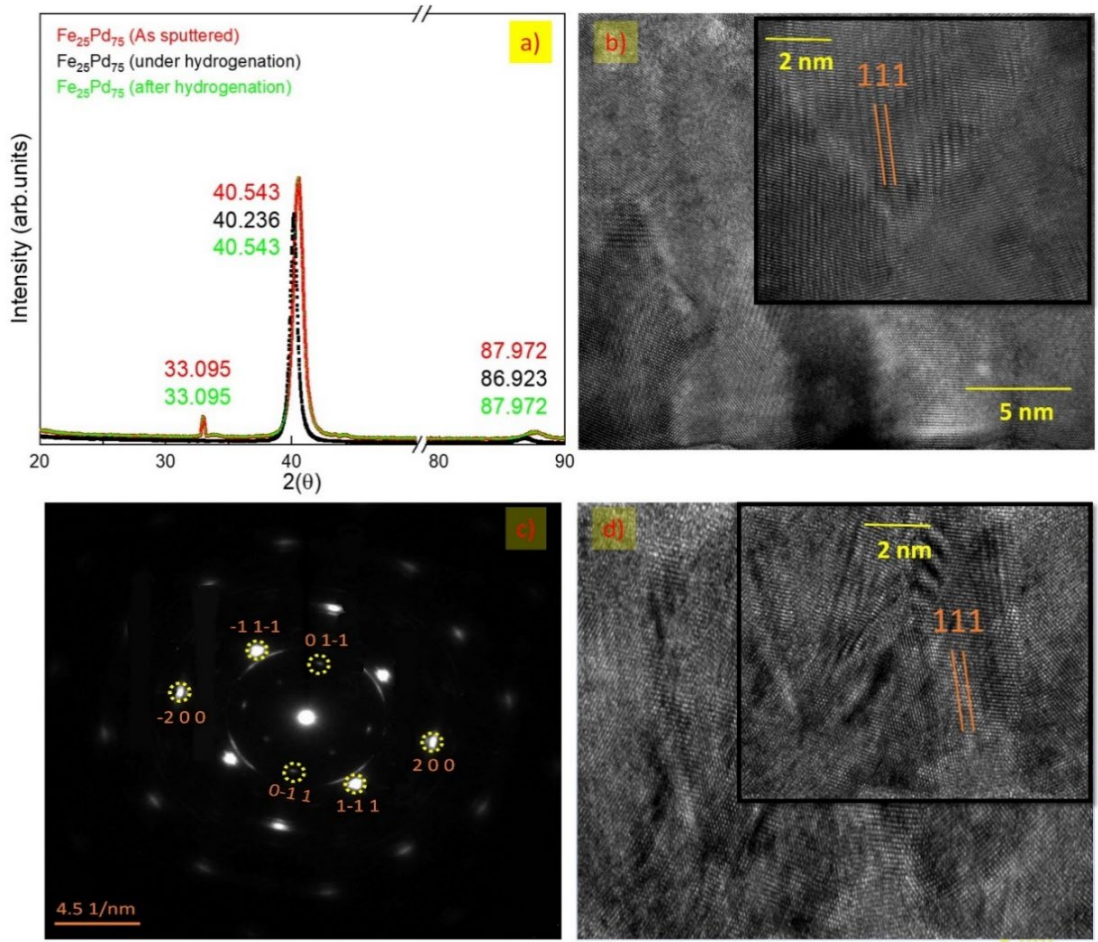

*Fig. 2: Structural characterization of the $Fe_{25}Pd_{75}$ alloy film: (a) Bragg-Brentano XRD spectra for the as-sputtered film in its virgin state (in red), hydrogenated state (in black), and de-hydrogenated state (in green). (b) High-Resolution Transmission Electron Microscopy (HRTEM) image of the film before hydrogenation, with the inset showing the growth direction in 111 lattice planes and d-spacing. (c) Selected Area Electron Diffraction (SAED) pattern of the pre-hydrogenated film. (d) HRTEM spectra of the $Fe_{25}Pd_{75}$ film after hydrogenation.*

*Table. 1: Parameters extracted from the XRD spectra (Fig. 2).*

| Measuring conditions | 2θ | d-spacing | FWHM |
|---|---|---|---|
| In $N_2$ | 40.25 | 2.248 | 0.56 |
|  | 86.96 | 1.123 | 1.02 |
| In $H_2$ | 40.11 | 2.249 | 0.56 |
|  | 86.62 | 1.124 | 0.715 |
| In $N_2$ (after $H_2$ exposure) | 40.25 | 2.248 | 0.56 |
|  | 86.96 | 1.123 | 1.02 |

These results suggest the presence of compressive strain in the in-plane direction upon hydrogenation, as discussed later. TEM analysis of the film's cross-section was performed to confirm the XRD findings. Fig. 2(b) and Fig. 2(d) display the High-Resolution Transmission Electron Microscopy (HRTEM) images of the $Fe_{25}Pd_{75}$ film in its as-deposited state and after prolonged exposure to $H_2$, respectively. The d-spacing of the (111) planes in the as-deposited film is smaller than that of the hydrogenated condition, which is consistent with the XRD observations. Subsequently, the Selected Area Diffraction (SAED) pattern of the as-deposited film, presented in Fig. 2(c), exhibits diffraction spots in a specific ring, indicating the [111] fiber texture present in the film's normal direction. Some diffraction spots belonging to the (200) plane can also be observed in the subsequent ring after the (111) plane. This reveals the polycrystalline nature of the film. Additionally, two innermost (110) diffraction spots are visible, suggesting the presence of pure Fe grains with a BCC structure. Our HRTEM data also indicates an irreversible reduction in grain boundaries following hydrogen exposure. For comparison, an HRTEM image of the virgin sample is provided in Fig. 2(b).

## *Dynamical magnetic characterization of the $Fe_{25}Pd_{75}$ film*

We employed FMR for the magnetic characterization of the film. In addition, we conducted magneto-optical Kerr effect measurements of the film to enable comparison with a previous study on FePd alloy films [22].

The inset to Fig. 3(a) displays typical differential FMR peaks $dP/dH_{DC}$ vs $H$ for the film, taken at 12 GHz while applying a constant magnetic field perpendicular to the film plane. This is referred to as out-of-plane (OOP) FMR. Similar traces were recorded at various frequencies (3 to 16 GHz) in the presence of nitrogen gas, known as the virgin state of the sample, during the sample's first exposure to 100% hydrogen, and after complete desorption of hydrogen gas when refilling the chamber with nitrogen. All measurements were taken at room temperature and atmospheric pressure.

To extract the material's magnetic parameters, the raw traces were fitted with the expression:[23]

$$\frac{dP}{dH_{DC}} = K1 \frac{4\Delta H(H - H_{res})}{[4(H - H_{res})^2 + (\Delta H^2)]^2} - K2 \frac{(\Delta H)^2 - 4(H - H_{res})^2}{[4(H - H_{res})^2 + (\Delta H)^2]^2} + slope H \quad 1$$
$$+ offset,$$

where $H$ is the applied magnetic field, $H_{res}$ is the resonance field, $\Delta H$ is the resonance linewidth, and *K1* and *K2* are the amplitudes of the symmetric and antisymmetric parts of the complex Lorentzian, respectively. The constants *slopeH* and *offset* account for the potential magnetic background signal in the experimental data. The extracted resonance fields $H_{res}$ were then plotted as a function of microwave frequency (symbols in Fig. 3) and fitted with the perpendicular-to-plane Kittel equation:

$$\omega = \gamma\left(H_0 - M_{eff}\right). \qquad 2$$

Here, $\omega = 2\pi f$ is the circular microwave frequency, at which the FMR measurements were recorded, $\gamma$ is the gyromagnetic ratio of the magnetic system, and $M_{eff}$ is the effective magnetization. The solid lines in Fig. 3 are fits to the experimental points. The black dots and lines represent the virgin sample, the red ones correspond to the hydrogenated state, and the green ones indicate the complete removal of hydrogen from the sample environment.

The excellent agreement between the model described by Eq. 2 and the experimental data enabled us to extract values of $\gamma/(2\pi)$ and $4\pi M_{eff}$ from the fits. For the virgin state of the sample, these were found to be 1.54 MHz/Oe and 4350 Oe, respectively. The parameter $4\pi M_{eff}$ represents the difference between the film's saturation magnetization $4\pi M_s$ and the effective field of perpendicular magnetic anisotropy (PMA) $H_u$. Unfortunately, FMR does not allow separating the two parameters. Thus, we supplemented the FMR measurements of the virgin state of the sample with Quantum Design MPMS-3 SQUID-VSM (VSM) measurements of $4\pi M_s$. The VSM measurements yielded a $4\pi M_s$ value of 5020 Oe. Hence, in the virgin state $H_u = 4\pi M_s - 4\pi M_{eff} = 410$ Oe. Thus, we may conclude that the film possesses a bulk uniaxial easy axis perpendicular magnetic anisotropy of moderate strength.

As shown in Table. 2: Magnetic parameters of the pristine and hydrogenated sample measured at 12 GHz., $4\pi M_{eff}$ increases by 400 Oe in the presence of hydrogen gas. Based on Eq. 2, this can be interpreted either as a decrease in $H_u$ or an increase in $4\pi M_s$ by the same amount, 400 Oe. However, a simultaneous change in both quantities is also possible, with $H_u$ decreasing and $4\pi M_s$ increasing, resulting in a net reduction in $4\pi M_{eff}$ by 400 Oe.

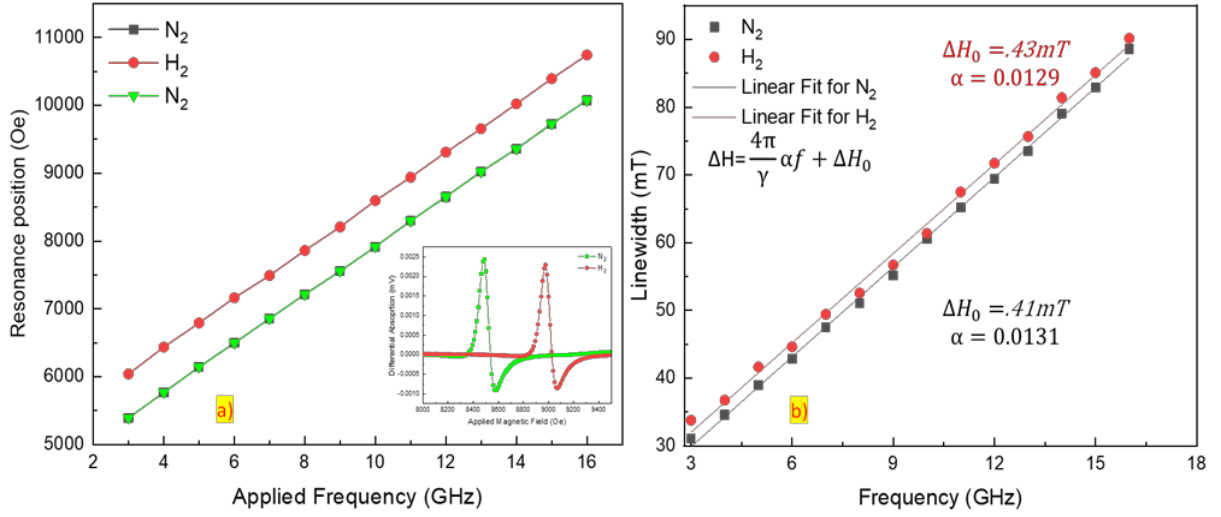

*Fig. 3: Microwave frequency versus FMR peak position dependence for pristine $Fe_{25}Pd_{75}$ alloy film (in green), hydrogenated $Fe_{25}Pd_{75}$ alloy film (in red), and dehydrogenated $Fe_{25}Pd_{75}$ alloy film (in black). The inset displays the FMR spectra of $Fe_{25}Pd_{75}$ film in $N_2$ and $H_2$ at 12 GHz. (b) Linear fit of the field linewidth (ΔH) as a function of the applied frequency.*

Fig. 3(b) displays the FMR peak width as a function of frequency *f*. The peak width characterizes magnetic losses in the material. It is observed that hydrogen absorption increases the FMR linewidth, leading to an increase in magnetic losses. Fitting the data with straight lines provides values of the Gilbert damping constant, $\alpha$, and inhomogeneous linewidth broadening ($\Delta H_0$). These values are shown in Fig. 4(b), along with the associated equation used to fit the data. The value of $\alpha$ remains largely unaffected by the absorption of hydrogen gas. However, there is a noticeable effect on $\Delta H_0$ – upon absorption of hydrogen gas $\Delta H_0$ increased from 4.1 Oe to 4.3 Oe. This change is reversible; upon desorption of hydrogen from the sample, the parameters return to their original values for the virgin state. This is a highly attractive feature for hydrogen gas sensing.

*Table. 2: Magnetic parameters of the pristine and hydrogenated sample measured at 12 GHz.*

| Sample | Linewidth (Oe) | Resonance Position (Oe) | Landé g-factor | $4\pi M_{eff}$ (Oe) |
|---|---|---|---|---|
| $Fe_{25}Pd_{75}$ continuous film | 69 | 8576 | 1.54 | 4610 |
| $Fe_{25}Pd_{75}$ continuous film ($H_2$) | 71 | 9105 | 1.55 | 5010 |
| $Fe_{25}Pd_{75}$ continuous film ($N_2$) | 69 | 8571 | 1.54 | 4600 |

Fig. 4 presents the results of the characterization using longitudinal MOKE. Fig. 4(a) shows the raw MOKE hysteresis loops. The main result of Fig. 4(a) displays the coercivity extracted from the hysteresis loops as a function of $H_2$ pressure and presented in Fig. 4(b). $H_2$ pressure varied in the range from 110 kPa to 220 kPa. Immediately after $H_2$ exposure, coercive field ($H_c$) drops to 233 Oe and then slightly decreases with an $H_2$ pressure of up to 220 kPa. The $H_c$ reduces by about 11.14% compared to the pristine $Fe_{25}Pd_{75}$ value in the first step of exposure to 105 kPa of hydrogen. The decrease of $H_c$ as a function of $H_2$ pressure could be mainly attributed to a decrease in magneto-crystalline anisotropy resulting from the absorption of $H_2$ [24, 25]. Moreover, the $H_2$ absorption led to an increase in the amplitude of the magneto-optical Kerr signal, which is consistent with other reports [22, 26].

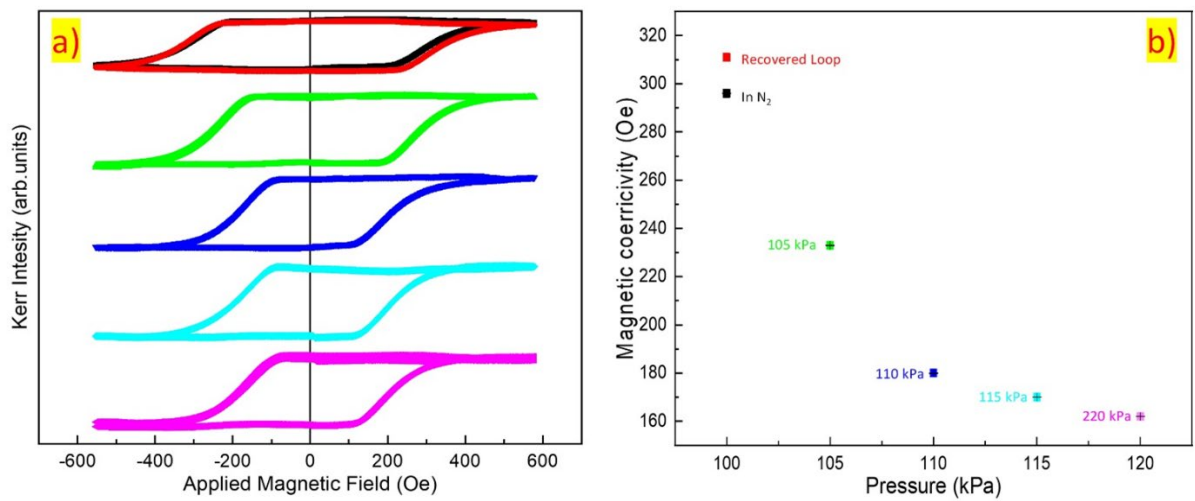

*Fig. 4: (a) Corresponding MOKE hysteresis loops of $Fe_{25}Pd_{75}$ sample. (b) Coercive field as a function of hydrogen pressure in an enclosed chamber.*

The observed changes in the magnetic properties of the FePd alloy thin films upon hydrogen incorporation can be explained by examining the strain states of the film before and after hydrogenation. As shown in Fig. 2(a), the (111) diffraction peaks of FePd shift to a lower 2θ angle upon hydrogenation. Since hydrogen incorporation can modify the crystalline structure and thus affect the magnetic properties of the material, including its coercivity and magneto-crystalline anisotropy. Fig. 4(a) illustrates the correlation between hydrogen pressure (or the degree of hydrogen incorporation) and the coercivity of the material, implying the presence of tensile lattice strain in the out-of-plane direction and compressive lattice strain in the in-plane direction. In a previous study, Feng et al. demonstrated that the $H_c$ of FePt thin films decreases by 80% with the introduction of a substrate-induced in-plane compressive strain of -2.18% [27]. A similar trend is observed in the current study: the $H_c$ of the FePd films decreases upon exposure to hydrogen. The corresponding changes in the in-plane hysteresis loop reported by Feng and colleagues [28] also display a trend consistent with the findings of the present study.

## DFT simulations

As demonstrated above, hydrogen gas absorption leads to an increase in $4\pi M_{eff}$ for the film. This may be interpreted as an increase in saturation magnetization or a decrease in the effective field of PMA. PMA at a FM-Pd interface is typically induced due to the hybridization of the electronic orbitals at the interface. This is a consequence of the difference in the electronic structure between the ferromagnetic material and Pd.

Furthermore, XRD analysis unveiled a notable augmentation in the film thickness when exposed to hydrogen gas. This increase in thickness implies a rise in the sample volume. If the magnetic moment of the film remains constant, the increase in thickness must result in an increase in saturation magnetization. Moreover, it is known that hydrogen absorption affects the electronic properties of metals. Their electronic band structure changes; consequently, the magnetic properties of alloys incorporating $H_2$-absorbing metals also change [29].

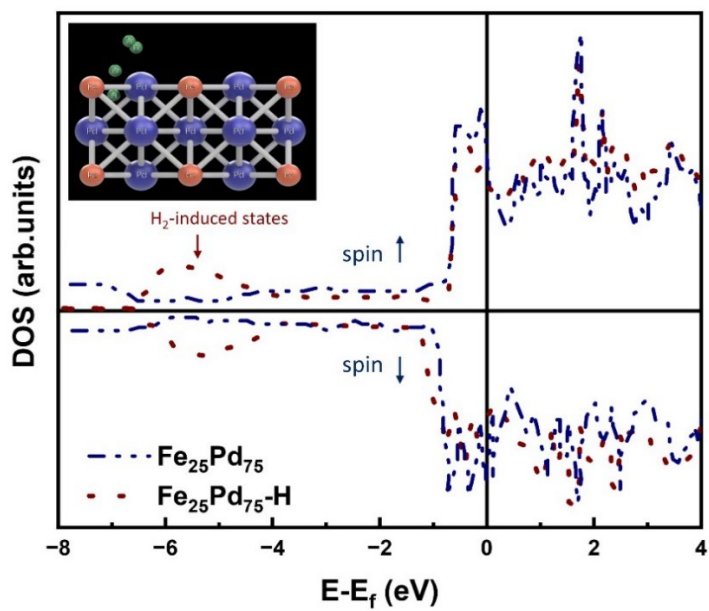

*Fig. 5: Calculated DOS profiles of 111 $Fe_{25}Pd_{75}$ and $Fe_{25}Pd_{75}$-H systems. The arrow in the figure indicates the hydrogen-induced states. The vertical line at E=0 shows the Fermi level.*

This may impact both saturation magnetization and PMA. For instance, Ref. [22] attributed the reduction in $H_c$ of Co/Pd multilayered films to changes in the electronic band structure of the material. Additionally, in pure bulk Pd, $H_2$ absorption leads to a decrease in Pd's magnetic susceptibility, which approaches zero as the d bands shift downward in Fermi energy due to increased bonding and antibonding of hydrogen 1s bands [29].

To understand how hydrogen absorption affects the electronic structure, we performed Density Functional Theory (DFT) simulations. The side view of the simulated $Fe_{25}Pd_{75}$-H system is shown in the inset of *Fig. 5*. The simulations revealed that the H-atoms prefer to sit close to Pd atoms at 1.49 Å from them. The corresponding binding energy value is −2.25 eV.

*Fig. 5* displays the calculated density of states (DOS) for the optimized $Fe_{25}Pd_{75}$ lattice grown in [111] direction and the respective $Fe_{25}Pd_{75}$-H system. The calculations reveal that $Fe_{25}Pd_{75}$ exhibits a metallic character, with its valence band dominated by d-orbitals of Fe and Pd. The conduction band consists of the d orbitals of iron. A significant contribution to the DOS by the d orbitals of Fe with energies between 0 and 2 eV results in magnetic behavior. The presence of a magnetic moment is observed as a considerable difference in DOS for spin up and spin down within this energy range. As per the Bruno's theory, the magnetic moments are coupled to the atomic lattice through the spin-orbit interaction, which results in a substantial anisotropic contribution. Also, the electrical structure of the neighboring atomic layers, the band structure, and the type of bonding at the interface all affect the strength of this interaction and the ensuing anisotropy [30]. However, it focuses on the atomic layers adjacent to the interface, potentially overlooking the influence of deeper layers or the bulk properties on the overall magnetic anisotropy. Table 3: presents values of the magnetic moment per atom extracted from this result. As shown in the table, the Fe atoms are characterized by an average magnetic moment of 0.72 $\mu_B$ per atom. Additionally, Pd also exhibits a small magnetic moment of 0.015 $\mu_B$ per atom. The magnetic polarization of Pd is a well-known phenomenon in Pd-rich magnetic alloys [31-37].

*Table 3: Calculated spin, orbital, and total magnetic moment per atom.*

| Fe$_{25}$Pd$_{75}$ | Spin magnetic moment (μ$_B$/atom) | Orbital magnetic moment (μ$_B$/atom) | Total magnetic moment (μ$_B$/atom) |
|---|---|---|---|
| Fe | 0.72 | 0.060 | 0.78 |
| Pd | 0.015 | 0.005 | 0.02 |
| H | - | - | - |
| Fe$_{25}$Pd$_{75}$-H | m$^s$ | m$^l$ | m$^t$ |
| Fe | 0.77 | 0.067 | 0.837 |
| Pd | 0.03 | 0.011 | 0.041 |
| H | - | - | - |

As shown in *Fig. 5*, the metallic character of Fe$_{25}$Pd$_{75}$ is preserved upon H adsorption. Similar to pristine Fe$_{25}$Pd$_{75}$, the valence band of Fe$_{25}$Pd$_{75}$-H is dominated by d-orbitals of Fe and Pd, while the conduction band consists of Fe's d orbitals. The hydrogen-induced energy bands can be observed at around -6 eV from the Fermi energy E=0. Their shape and energy closely agree with those of pure Pd reported in the literature [38, 39]. Consequently, from the electronic structure perspective, hydrogen absorption induces similar changes to the Fe$_{25}$Pd$_{75}$ band structure as seen in the case of pure Pd.

The total magnetic moment of Fe$_{25}$Pd$_{75}$-H increases to 0.8 μ$_B$, with contributions of 0.78 μ$_B$ and 0.03 μ$_B$ per Fe and Pd atoms, respectively. As the magnetic polarization of Pd plays a crucial role in magneto-elastic coupling, the increase in magnetostriction might be significant, leading to a slight change in bulk-PMA energy [31]. We also hypothesize that both changes - a decrease in $H_u$ or an increase in $4\pi M_s$ - could theoretically result in an increase in $M_{eff}$. However, if the magnetic moment and thus the $M_s$ are directly increasing due to the influence of H$_2$, it seems more plausible to assume that the increase in $M_{eff}$ is largely due to an increase in $4\pi M_s$.

## Prototype of an FMR-based integrated hydrogen gas sensor

In reference [40], we proposed implementing an FMR-based hydrogen gas sensor as an integrated, lithographically formed device. The concept involved depositing an FePd alloy film on top of a central conductor (called "signal line") of a miniature microwave coplanar transmission line (CPL). The CPL acts similarly to the microstrip line used in the microstrip line FMR method. Also, we preferred a CPL over a microstrip line due to several factors. The primary rationale for selecting a CPL is that, for the microstrip line geometry, the strip width and substrate thickness are not arbitrary and need to be specific to attain a characteristic impedance (Zc) of 50 Ohms.

Let us consider a sapphire substrate, characterized by a dielectric constant (epsilon) of 11. To achieve Zc=50 Ohms with this substrate, the thickness must be equal to the strip width. Therefore, for a 50-micron strip, we would need a 50-micron thick substrate. To gradually decrease the strip width to microscopic sizes while maintaining the same Zc, we would need to correspondingly alter the substrate thickness. This method proves to be quite impractical. Contrastingly, CPL doesn't have this constraint as its Zc depends exclusively on the width of the signal and ground lines, the gap between them, and the dielectric constant of the substrate. This attribute allows us to gradually decrease CPW dimensions while keeping the Zc constant.

Also, reducing the width of the microstrip line enhances the FMR absorption signal, or the height of the FMR peak (see, e.g. Fig. 3 in Ref. [41]). However, if the width is too small, ohmic losses in the transmission line become significant, causing the FMR absorption signal to decrease. This trade-off results in an optimum microstrip line width of tens of micrometers. Technically, CPL is a more suitable geometry for a miniature (micron-scale) stripline microwave waveguide than a microstrip line. In this case, the magnetic film can be directly deposited on top of the CPL's signal line to produce an FMR response. To further increase the FMR absorption signal, the film's length along the CPL can also be increased, although this will raise the ohmic losses of the CPL.

In this study, we investigated the same $Fe_{25}Pd_{75}$ film as described in previous sections, but while deposited on the surface of an Au-Al alloy-based CPL signal line. A detailed fabrication process is illustrated in Fig. 6(a). The film forms a rectangle sitting on top of a 50 μm wide signal line of a CPL Fig. 6(b). The dimensions of the resultant waveguide are demonstrated in Fig. 6(c). The rectangle width is also 50 μm, such that its edges align with the signal line's edges, and the rectangle length along the signal line is 4 mm. We also fabricated a continuous reference film on a Si substrate in the same FePd film deposition run. FMR measurements of the sample on top of the CPL were conducted in the in-plane FMR configuration. This arrangement was chosen for a technical reason. We used microscopic microwave probes "Picoprobes" from GGB Industries to contact the microscopic CPL. Our picoprobe station allows the application of a magnetic field in the sample plane only, hence the choice of in-plane magnetization direction. Accordingly, the FMR traces for the reference continuous film were also taken in-plane.

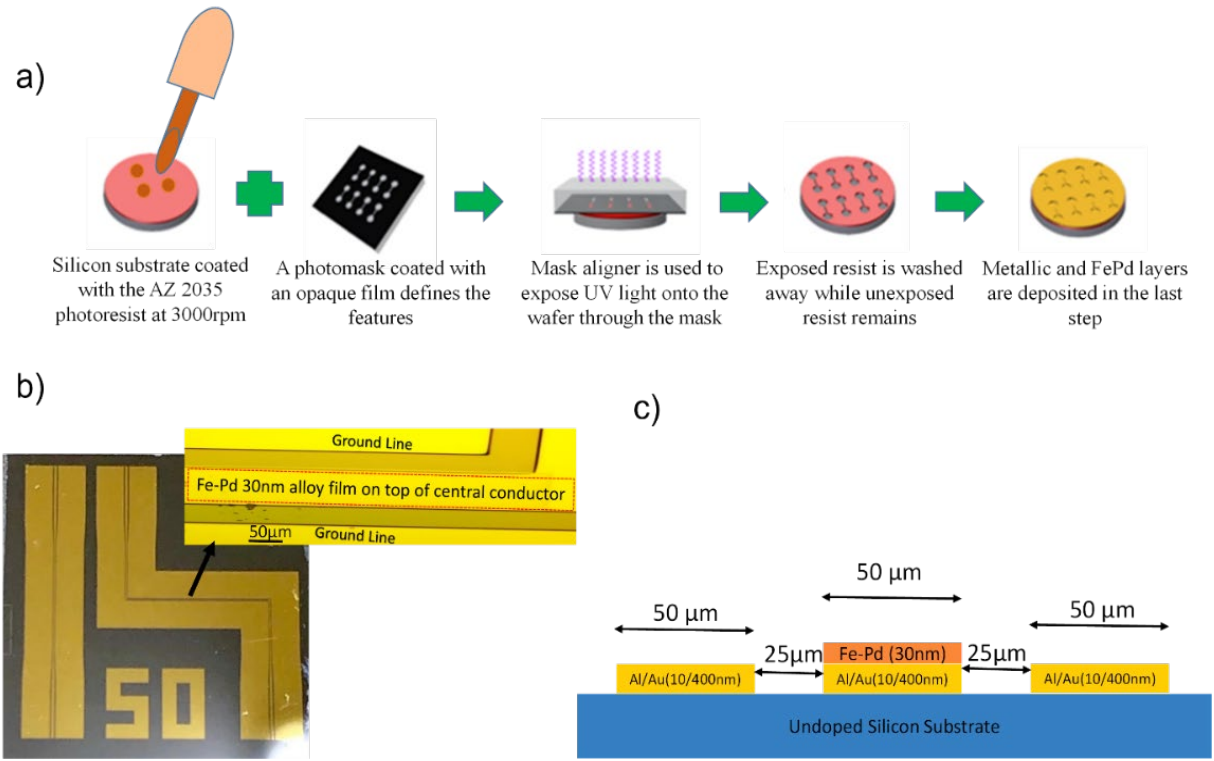

*Fig. 6 illustrates the process of coplanar transmission line patterning utilizing photolithography and lift-off techniques. Panel (a) depicts the sequential steps in this procedure, which uniquely features lithographic operations carried out prior to the deposition of FePd, as shown in panel (b). The dimensions of the resultant waveguide are demonstrated in panel (c).*

Fig. 7(a) displays the FMR absorption peaks for both samples. The absorption signal is so strong (1.5 mV) that we do not need to use a lock-in amplifier and a field-modulated FMR to register it. Instead, this can be done using a digital voltmeter. Consequently, the absorption peak shape appears as a Lorentzian. Fig. 7(a) reveals that the resonance linewidth is larger for the $Fe_{25}Pd_{75}$ based CPL film grown on the Au-Al film compared to the reference film. Panel (b) of Fig. 7 demonstrates the effect of hydrogen gas on the film on the CPL. The measurements were taken in an atmosphere containing 3% hydrogen gas and 97% nitrogen as a carrier gas. The inset to the panel displays the registered FMR traces. The main field of the panel shows the hydrogen induced FMR peak shift as a function of microwave frequency. Table 4: presents the sample's magnetic parameters extracted from the fits of the FMR traces. The data show that just 3% of hydrogen results in an FMR peak shift of 94 Oe, which is almost half of the linewidth.

*Table 4: The Magnetic parameters of the pristine and hydrogenated sample. The data is extracted from Fig. 7(a)*

| Sample | Linewidth (Oe) | Resonance Position (Oe) | Landé $g$-factor | $4\pi M_{eff}$ (Oe) |
|---|---|---|---|---|
| Virgin $Fe_{25}Pd_{75}$ CPL | 210 | 1001 | 1.55 | 4440 |
| Hydrogenated $Fe_{25}Pd_{75}$ CPL | 220 | 1095 | 1.54 | 5210 |
| de-hydrogenated $Fe_{25}Pd_{75}$ CPL | 210 | 1001 | 1.55 | 4430 |

Lastly, we investigated the time dependence of the FMR response by rapidly introducing or removing hydrogen gas to or from the gas chamber that houses the pico-probe station. The magnetic field applied to the CPL was fixed at 1001 Oe, corresponding to the minimum of the FMR absorption trace for the nitrogen atmosphere. We cycled several times between 100% nitrogen gas and 3% hydrogen in 97% nitrogen and recorded the amplitude of the FMR absorption signal as a function of time.

The results shown in Fig. 8. indicate that $Fe_{25}Pd_{75}$ alloy systems exhibit rapid and highly reversible responses to hydrogen gas absorption and desorption. With just 3% of hydrogen gas concentration in the environment, it takes about 30 seconds for the $H_2$ absorption process to complete and 67 seconds for the gas desorption process. These response times are the fastest ever reported for $Fe_{25}Pd_{75}$ alloy systems, highlighting the potential of this material for hydrogen gas sensing applications.

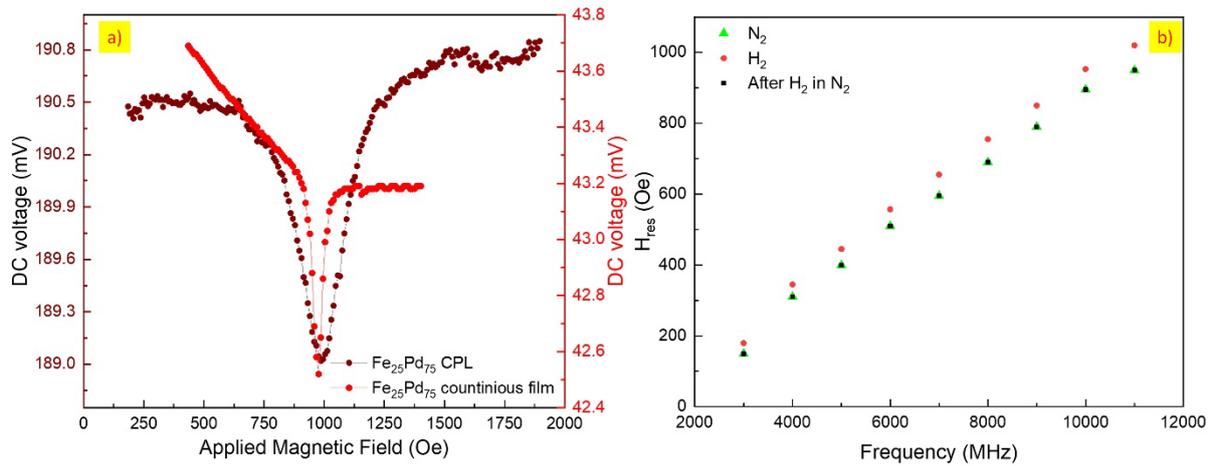

*Fig. 7: Presents a linewidth-adjusted comparison of the $Fe_{25}Pd_{75}$ CPL and $Fe_{25}Pd_{75}$ continuous thin film at 10 GHz. (a) The FMR response of the continuous film and the film on top of a CPL is displayed in red and wine, respectively. Panel (b) illustrates the resonance position versus frequency graph in an in-plane FMR configuration.*

In addition to the fast response times, the high reversibility of the system is noteworthy. Cycling between hydrogen and nitrogen gas environments multiple times resulted in responses that were practically the same for each cycle, demonstrating the stability and reliability of the $Fe_{25}Pd_{75}$ alloy system as a potential hydrogen gas sensor. This reversibility is essential for practical applications, as it ensures the consistency and accuracy of the sensor's performance over multiple cycles.

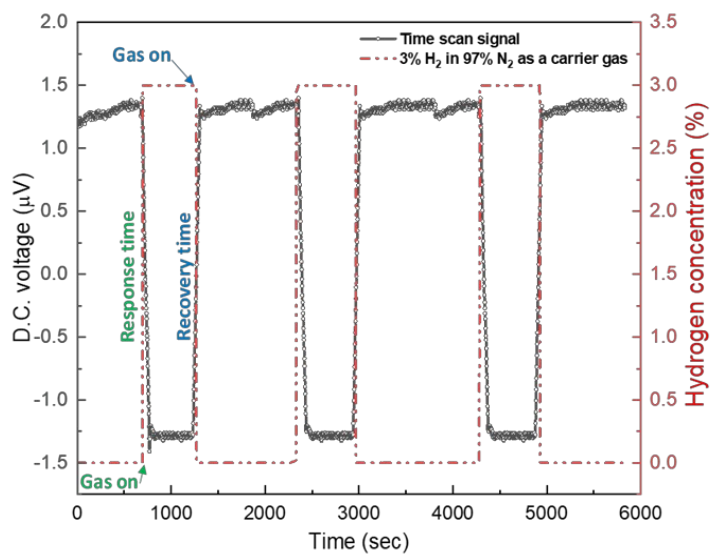

*Fig. 8: Shows the time-resolved FMR for the $Fe_{25}Pd_{75}$ device is shown as the sample is cycled between pure nitrogen and a mixture of 3% hydrogen gas in nitrogen. The red dotted line indicates the concentration of hydrogen gas, plotted on the right-hand y-axis.*

Furthermore, broadening of the FMR linewidth in polycrystalline materials generally occurs due to factors such as inhomogeneities, different magnetic phases, non-magnetic phase inclusions, stress, porosity, and anisotropy [28, 42, 43]. In single-crystal materials, anisotropy, and the presence of ions with spin-orbit coupling would widen the linewidth. While it is difficult to pinpoint the exact reasons for linewidth broadening in your system, it is speculated that the broadening could be caused by increased stress in the presence of $H_2$.

The hydrogen-induced FMR peak shift observed in the present work is larger in Pd-rich ($Fe_{25}Pd_{75}$) samples than in Fe-rich samples. One possible explanation is that the larger Pd concentration in the Fe matrix enhances the amount of $H_2$ absorption, leading to a greater hydrogen-induced magnetization change. This aligns with studies on other metallic thin films that absorb $H_2$, in which changes in their magnetic properties are also believed to result from electron transfer. In pure bulk Pd, $H_2$ absorption leads to a decrease in magnetic susceptibility, which goes to zero as the d bands shift downward in energy relative to the Fermi energy due to increased bonding and antibonding of the hydrogen 1s bands [29]. Pd is known to be paramagnetic with high magnetic susceptibility and can undergo spontaneous spin polarization when close to ferromagnetic materials [29]. In bulk FePd alloys, $H_2$ absorption leads to a smaller magnetic moment. For example, in CoPd films, Pd atoms become polarized near Co atoms, resulting in the magnetization of the Pd layers. Another possibility is mentioned in Ref. [44], where they found that in FePd alloys, the long-range R.K.K.Y. coupling between Fe atoms is significantly reduced upon $H_2$ absorption.

# Conclusion

In conclusion, this study investigates the effects of $H_2$ on the magnetic, structural, and electronic properties of the FCC phased $Fe_{25}Pd_{75}$ alloy system, with potential applications in magnonic-based $H_2$ gas sensing. Significant $H_2$-induced FMR peak shifts were observed in the Pd-rich $Fe_{25}Pd_{75}$ film, and these shifts were later explained through both theoretical and experimental approaches. We speculated that the alterations in both $H_u$ and $4\pi M_s$ could lead to an increase in $M_{eff}$. However, it was found that if the magnetic moment and $M_s$ increase directly due to the influence of $H_2$, it is more likely that the increase in $M_{eff}$ is primarily attributable to an increase in $4\pi M_s$.

Furthermore, the study demonstrated that the amplitude of the FMR signal increases when driven by a coplanar transmission line. The large amplitude of the FMR signal facilitated direct measurements using a simple microwave diode and a digital voltmeter, streamlining the data collection process. Overall, the results provide valuable insights into the behavior of the $Fe_{25}Pd_{75}$ alloy system under $H_2$ exposure, paving the way for future development of magnonic-based $H_2$ gas sensors.

## Declaration of Competing Interest

The authors declare that they have no known competing financial interests or personal relationships that could have appeared to influence the work reported in this paper.

## Acknowledgement

This work is supported by the Research Training Program (RTP) Scholarship from the University of Western Australia (UWA). The authors acknowledge the facilities, as well as the scientific and technical assistance, of the Australian National Fabrication Facility (ANFF) at the University of Western Australia. We also express our gratitude to Professor Martin Saunders for his expertise in Transmission Electron Microscopy (TEM) analysis at the Centre for Microscopy, Characterisation and Analysis (CMCA) at UWA. Currently, Shahbaz holds the position of a CSIRO Early Research Career (CERC) Postdoctoral Fellow in the CSIRO Environment, with funding provided by the Autonomous Future Science Platform (AS-FSP).